\documentclass[preprint,5pt]{elsarticle}
\usepackage[T1]{fontenc}
\usepackage[latin9]{inputenc}
\usepackage{amsmath}
\usepackage{graphicx}
\usepackage{amssymb}
\usepackage{esint}
\usepackage{amsthm}
\usepackage{dcolumn}
\usepackage{bm}
\usepackage{multirow}
\journal{Physical Review A }
\begin{document}
\newtheorem{conjecture}{Conjecture}\newtheorem{corollary}{Corollary}\newtheorem{theorem}{Theorem}
\newtheorem{lemma}{Lemma}\newtheorem{observation}{Observation}\newtheorem{definition}{Definition}
\newtheorem{remark}{Remark}\global\long\global\long\def\ket#1{|#1 \rangle}
 \global\long\global\long\def\bra#1{\langle#1|}
 \global\long\global\long\def\proj#1{\ket{#1}\bra{#1}}
\begin{frontmatter}

\title{Optimal Quantum State Tomography via Weak Value}

\author[gc1,gc2]{Xuanmin Zhu\corref{cor}}
\ead{zhuxuanmin2006@163.com}
\author[gc1]{Dezheng Zhang}
\author[gc1]{Runping Gao}
\author[xd1]{Qun wei}
\author[xd2]{Lixia Liu}
\author[sd1]{Zijiang Luo}
\address[gc1]{Center for Quantum Information, School of Information, Guizhou University of Finance and Economics, Guiyang, Guizhou 550025, China}
\address[gc2]{International Joint Research Center for Data Science and High-Performance Computing, School of Information, Guizhou University of Finance and Economics, Guiyang, Guizhou 550025, China}
\address[xd1]{School of Physics and Optoelectronic Engineering, Xidian University, Xi'an 710071, China}
\address[xd2]{School of Mathematics and Statistics, Xidian University, Xi'an 710071, China}
\address[sd1]{Institute of Intelligent Manufacturing, Shunde Polytechnic, Guangdong Shunde 528300,China}
\cortext[cor]{Corresponding author. Tel.: +86 15802914790.}
\begin{abstract}
To improve the efficiency of the state tomography strategy via weak value, we have searched the optimal coupling strength between the system and measuring device. For an arbitrary d-dimensional quantum system, the optimal strengths being used in measuring the real and imaginary parts of the density matrix are obtained. The optimal efficiency of the state tomography has also been studied by using mean square error. The minimal mean square errors in the reconstructed density matrices have been derived. The state tomography strategy studied in this article may be useful in the measurement of the unknown quantum states.
\end{abstract}

\begin{keyword}
Qauantum state tomography \sep Weak measurement \sep Weak value

\end{keyword}

\end{frontmatter}

\section{Introduction}\label{sec1}
An unknown state of the quantum systems could be reconstructed by using quantum state tomography (QST) with perfect accuracy in theory ~\cite{niel,qst1,qst2,qst3,qst4,qst5,qst6,qst7,qst8, qst9, qst10, qst11, qst12, qst13, qst14,qst15,qst16,qst17,qst18,qst19,qst20}. In practice, the reconstruction of the quantum state suffers from random errors, which can be reduced by repeated measurements. Roughly speaking, the random error decreases as the number of the measurements. The number of the measurements is always determined by the identical quantum systems. Then, the identical unknown quantum systems are the important physical resources in QST. The optimization problem in QST is how to decrease the random errors by measuring a finite number of quantum systems.

Quantum  state tomography via weak value (QSTW) is a strategy in which the elements of the unknown density matrix can be directly reconstructed by the measurement results of the weak values~\cite{aav, wk1, wk2, wk3, wk4, dsm1, dsm2,dsm3,dsm4,dsm5}. The scheme of QSTW is valid for an arbitrary d-dimensional unknown state and much more easily realized. Thus, much attention has been focused on the method of QSTW~\cite{dsm6,dsm7,dsm8,dsm9,dsm10,dsm11,dsm12,dsm13,dsm14,dsm15,dsm16,dsm17,
dsm18, budi, yang}.

The coupling strength between the system and measuring device in QSTW is neither strong nor weak. And the real and imaginary parts of the density matrix are individually reconstructed. The coupling strengths of measuring the two parts of the matrix can be different. In this article, we derive the optimal measurement strengths of QSTW and the mean minimal square error of the reconstructed state in theory. Furthermore, we improve the efficiency by the Hermiticity of the reconstructed density matrix. The results of the Monte Carlo simulations show that the theoretical values are consistent with the simulated results.

The rest of this paper is organized as follows. The quantum state tomography via weak value is simply reviewed in Sec.~\ref{sec2}, and the optimal measurement strengths and the efficiency are derived in Sec.~\ref{sec3}. In Sec.~\ref{sec4}, we compare the efficiency of the optimal QSTW with that of the conventional QST. A short conclusion is given in Sec.~\ref{sec5}.

\section{Quantum state tomography via weak value}\label{sec2}
In this section, we simply review the quantum state tomography via weak value (QSTW) which is based on the weak measurement theory. We should point out that QSTW is the one named MDST (modified direct state tomography) in Ref.~\cite{zhu1}. The procedure of QSTW is performed by the following steps.

First, the quantum systems with the same unknown state $\rho_s$ are divided into $2d$ equal parts when the dimension of the systems is $d$. We perform a weak measurement on each part, the interaction Hamiltonian between the
system and measuring device can be expressed as
\begin{equation}\label{e1}
H_n=g\delta(t-t_0)A_{n}\otimes \sigma_x,
\end{equation}
where $\{A_n=|a_n\rangle\langle a_n|\}$ are the observables of the system, $\{|a_n\rangle\}$ is a basis of the system, and $g$ is the measurement strength. Without loss of generality, the initial state of the measuring device is $\rho_d = |0\rangle_d\langle 0|$,
where $|0\rangle_d$ is the eigenstate of the usual Pauli matrix $\sigma_z$ with the eigenvalue $1$. After the interaction the combine state of the system and measuring device is
\begin{equation}\label{e2}
\rho_n=U_n\rho_s\otimes\rho_dU_n^{\dagger},
\end{equation}
where $U_n=e^{-ig |a_n\rangle\langle a_n|\otimes \sigma_x}$, with $\hbar=1$.

Second, a strong projective measurement is implemented along a basis $\{|\psi_j\rangle\}$ on the quantum system. In QSTW, the bases $\{|a_n\rangle\}$ and $\{|\psi_j\rangle\}$ are chosen as the mutually unbiased bases (MUBs) which satisfy $\langle \psi_j|a_n\rangle=e^{2\pi jn i/d}/\sqrt{d}$~\cite{mub}. Conditioned on obtaining the system state $|\psi_j\rangle$ in the projective measurement, the final measuring device state is
\begin{equation}\label{e3}
\rho_d^{nj}=\frac{\langle \psi_j|U_n\rho_s\otimes\rho_dU_n^{\dagger}|\psi_j\rangle}{P_j},
\end{equation}
where $P_j$ is the probability of obtaining $|\psi_j\rangle$.

Third, the \emph{weak value} $W_{nj}$ which is defined as~\cite{dsm3}
\begin{equation}\label{e4}
W_{nj}=\frac{\langle \psi_j|a_n\rangle\langle a_n |\rho_s|\psi_j\rangle}{P_j}
\end{equation}
can determined by measuring the two observables $\sigma_R$ and $\sigma_I$ of the measuring device~\cite{zhu1, zhan, vall, gro, zou, zhu2, zhu3}
\begin{equation}\label{e5}
W_{nj}=\frac{1}{2g}\left[-\mathbf{Tr}(\rho_d^{nj} \sigma_R) +i\mathbf{Tr}(\rho_d^{nj}\sigma_I)\right],
\end{equation}
where
\begin{equation}\label{e6}
\begin{split}
\sigma_R &=\frac{g}{\sin g}\left[\sigma_y-\tan \left(\frac{g}{2}\right)\left( I-\sigma_z \right)\right],\\
\sigma_I &=\frac{g}{\sin g}\sigma_x.
\end{split}
\end{equation}

Finally, the matrix elements $\{\rho_{nm}\}$ of the unknown state $\rho_s$ can be reconstructed by the results of the weak values,
\begin{equation}\label{e07}
\langle a_n|\rho_s|a_m\rangle=\sum_{j}P_j\frac{\langle \psi_j|a_m\rangle}{\langle \psi_j|a_n\rangle} W_{nj}.
\end{equation}

Eq. (\ref{e6}) shows that the real and imaginary parts of the density matrix are measured independently. Thus the coupling strengths used to measuring the real and imaginary parts can be different, which are denoted as $g_R$ and $g_I$ respectively. In the next section we will derive the optimal $g_R$ and $g_I$.

\section{The optimal coupling strengths and the minimal error}\label{sec3}

In this section we search the optimal coupling strengths and the minimal error in the reconstructed state $\rho_r$. The mean-square error (MSE) is used to measure the discrepancy between the true state $\rho_s$ and the reconstructed state $\rho_r$~\cite{qst11,qst12,qst13}, defined as
\begin{equation}\begin{split}\label{e08}
\mathcal{E} \equiv E (\Vert \rho_r- \rho_s\Vert^2_{HS}),
\end{split}\end{equation}
where $\Vert \rho_r-\rho_s \Vert^2_{HS}$ is the Hilbert-Schmidt norm. We have
\begin{equation}\begin{split}\label{e09}
\mathcal{E}&=E(\mathrm{tr}[(\rho_r-\rho_s)^{\dagger}(\rho_r-\rho_s)])\\
&=\sum_{nm}E(|\rho_{nm,r}-\rho_{nm,s}|^2).
\end{split}\end{equation}
Thus MSE is the sum of the variances of all the elements of the density matrix. There is no system error in QSTW, we have $\rho_s=E(\hat{\rho}_r)$, where $\hat{\rho}_r$ is the estimator. From Eq. (\ref{e09}), the MSE can be expressed as
\begin{equation}\label{e10}
\mathcal{E}=\frac{1}{N}[\mathrm{tr}E(\hat{\rho}_r^\dagger\hat{\rho}_r)-
\mathrm{tr}(\rho_s^2)],
\end{equation}
where $N$ is the number of the measurements.

By Eq. (\ref{e07}), the estimator can be express as
\begin{equation}\label{e11}
\hat{\rho}_r=\sum_{nj}\frac{P_j W_{nj}}{\langle \psi_j|a_n\rangle}|a_n\rangle\langle \psi_j|.
\end{equation}
Using this equation and $|\langle \psi_j|a_n\rangle|=1/\sqrt{d}$, we have
\begin{equation}\begin{split}\label{e12}
\mathrm{tr}E(\hat{\rho}_r^\dagger\hat{\rho}_r)&=d\sum_{nj}E(P_jW_{nj}^*W_{nj}),\\
&=d\sum_{nj}E(Re(P_jW_{nj}^2))+E(Im(P_jW_{nj}^2)).
\end{split}\end{equation}
Follows from Eqs. (\ref{e4}) and (\ref{e5}),
\begin{equation}\begin{split}\label{e13}
E(Re(P_jW_{nj}^2))&=\sum_{n}\mathrm{tr}(\rho_{n,d}\sigma_R^2) \\
E(Im(P_jW_{nj}^2))&=\sum_{n}\mathrm{tr}(\rho_{n,d}\sigma_I^2),
\end{split}\end{equation}
where $\rho_{n,d}=\mathrm{tr}(U_n\rho_s\otimes \rho_d U_n^\dagger)$, the identity operator $\sum_j|\psi_j\rangle\langle\psi_j|=I_s$ is used in the derivation.
The expression of the density matrix $\rho_{n,d}$ is
\begin{equation}\begin{split}\label{e14}
\rho_{n,d}&=|0\rangle_d\langle 0|+\rho_{nn}[\sin^2 g(-|0\rangle_d\langle 0|+|1\rangle_d\langle 1|) \\
&+i\cos g \sin g(|0\rangle_d\langle 1|+|1\rangle_d\langle 0|),
\end{split}\end{equation}
where $\rho_{nn}=\langle a_n|\rho_s|a_n\rangle$ is the diagonal element of the density matrix $\rho_s$. Substituting Eqs. (\ref{e6}) and (\ref{e14}) into Eq. (\ref{e13}), by using the equation $\sum_n \rho_{nn}=1$, we obtain
\begin{equation}\begin{split}\label{e15}
E(Re(P_jW_{nj}^2))&=\frac{1}{4}\left(\frac{d}{\sin^2g_R}+\frac{2}{\cos^2(g_R/2)}\right) \\
E(Im(P_jW_{nj}^2))&=\frac{d}{4\sin^2g_I}.
\end{split}\end{equation}
From Eqs.(\ref{e10}), (\ref{e12}) and (\ref{e15}), we have the MSE of the reconstructed state
\begin{equation}\label{e16}
\mathcal{E}=\frac{1}{N}[\frac{d^2}{4}(\frac{1}{\sin^2g_R}+\frac{1}{\sin^2g_I})
+\frac{d}{2\cos^2(g_R/2)}-\mathrm{tr}(\rho_s^2)].
\end{equation}
The MSE $\mathcal{E}$ is not only dependent on the dimension of the system, but also the measurement strengths $g_R$ and $g_I$. We can search the optimal $g_R$ and $g_I$ to attain the minimal value of $\mathcal{E}$. By simple calculations, we obtain optimal measurement strengths
\begin{equation}\begin{split}\label{e17}
g_{R,o}&=\arccos\left(1+\frac{d}{4}-\sqrt{\frac{d}{2}+\frac{d^2}{16}}\right) \\
g_{I,o}&=\frac{\pi}{2}.
\end{split}\end{equation}
And the optimal MSE is
\begin{equation}\label{e18}
\mathcal{E}_o=\frac{1}{N}[\frac{3d^2}{8}+\frac{d}{2}\left(\sqrt{\frac{d}{2}+\frac{d^2}{16}}
+1 \right)-\mathrm{tr}(\rho_s^2)].
\end{equation}

We have performed the Monte Carlo simulations to verify our results. In the simulations, $10^3$ identical five-dimensional quantum systems are used to accomplish $10^2$ measurements. All the values of MSEs in Fig. 1 are averaged over $10^3$ repeated simulations to decrease the influence of the statistical fluctuations. As shown in Fig.1, the simulation results are consistent with $\mathcal{E}_o=[16.91-\mathrm{tr}(\rho_s^2)]/100$ given by Eq. (\ref{e16}). The optimal coupling strengths are $g_{R,o}=1.33$ and $g_{I,o}=\pi/2$.
\begin{figure}[t]
\centering \includegraphics[scale=0.5]{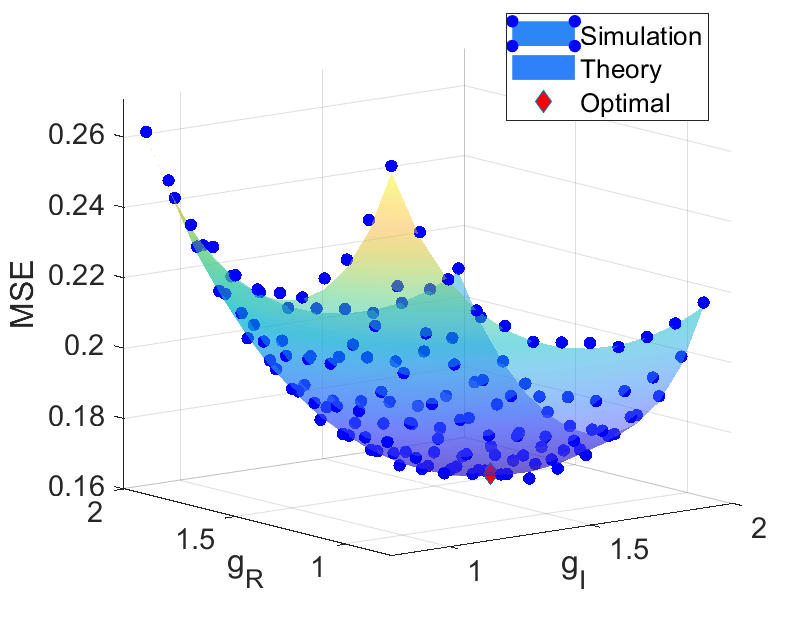} \caption{(Color online) The MSEs of the simulation results and the ones given by Eq. (\ref{e16}) in theory. The diamond represents the minimal MSE when $g_{R,o}=1.33$ and $g_{I,o}=\pi/2$.}
\label{fig1}
\end{figure}

\section{Hermitian density matrix and comparison}\label{sec4}
The reconstructed density matrix given by Eq. (\ref{e07}) is not a Hermitian matrix. By the Hermiticity of the density matrix $\rho_s$, we can obtain a Hermitian density matrix according to the formula
\begin{equation}\label{e19}
\hat{\rho}_r'=\frac{\hat{\rho}_r+\hat{\rho}_r^\dagger}{2}.
\end{equation}
As the elements of the density $\rho_{nm}$ and $\rho_{mn}$ given by Eq. (\ref{e07}) are measured independently when $n\neq m$. The number of the measurements of the $\rho_{nm}'=(\rho_{nm}+\rho_{mn}^\dagger)/2$ will be $2N$ if the number of the measurements of the $\rho_{nm}$ is $N$. The proportion of the off-diagonal elements in the density matrix is $(d-1)/d$, Then the MSE of the  off-diagonal elements of the $\hat{\rho}_r'$ is
\begin{equation}\label{e020}
\mathcal{E}_{off}= \frac{d-1}{2Nd}[\frac{d^2}{4}(\frac{1}{\sin^2g_R}+\frac{1}{\sin^2g_I})
+\frac{d}{2\cos^2(g_R/2)}-\mathrm{tr}(\rho_s^2)].
\end{equation}
The number of the measurements of the diagonal elements of $\hat{\rho}_r'$ remains unchanged $N$. However, the MSE of the imaginary part of the diagonal elements is eliminated, because $\hat{\rho}_r'$ is a Hermitian matrix. As the proportion of the diagonal elements is $1/d$, the  MSE of the diagonal elements of the $\hat{\rho}_r'$ is only the MSE of the real part, which can be expressed as
\begin{equation}\label{e021}
\mathcal{E}_{dia}= \frac{1}{Nd}[\frac{d^2}{4\sin^2g_R}
+\frac{d}{2\cos^2(g_R/2)}-\mathrm{tr}((Re\rho_s)^2)].
\end{equation}

Then the total MSE of the Hermitian density matrix $\rho_r'$ is
\begin{equation}\begin{split}\label{e20}
\mathcal{E}'&=\mathcal{E}_{off}+\mathcal{E}_{dia}\\ &=\frac{d+1}{2dN}[\frac{d^2}{4\sin^2g_R}
+\frac{d}{2\cos^2(g_R/2)}-\mathrm{tr}((Re\rho_s)^2)]\\
&+\frac{d-1}{2dN}[\frac{d^2}{4\sin^2g_I}-\mathrm{tr}((Im\rho_s)^2)].
\end{split}\end{equation}
As shown in Fig. 2, the results of the Monte Carlo simulations are consistent with the ones given by Eq. (\ref{e20}).

\begin{figure}[t]
\centering \includegraphics[scale=0.5]{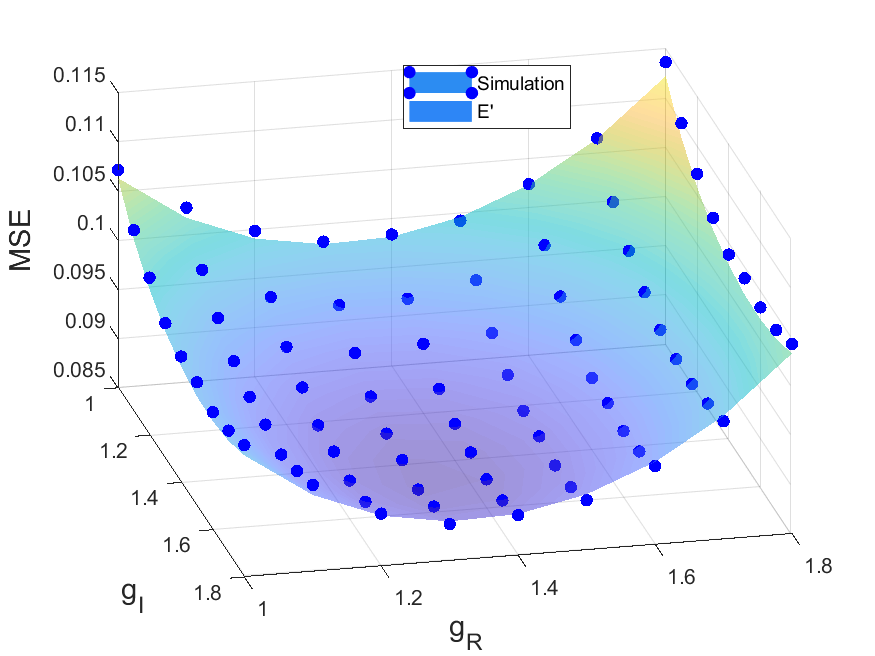} \caption{(Color online) The MSEs of the simulation results and the ones given by Eq. (\ref{e20}) in theory. In the simulations, the number of the measurements for the unknown five-dimensional quantum systems is $10^2$ . All the values of MSEs are averaged over $10^4$ repeated simulations to decrease the influence of the statistical fluctuations.}
\label{fig2}
\end{figure}
The optimal MSE of the density $\rho_r'$ is
\begin{equation}\begin{split}\label{e200}
\mathcal{E}_o' &=\frac{d+1}{2dN}[\frac{d^2}{8}
+\frac{d}{2}\left(\sqrt{\frac{d}{2}+\frac{d^2}{16}}+1\right)-\mathrm{tr}((Re\rho_s)^2)]\\
&+\frac{d-1}{2dN}[\frac{d^2}{4}-\mathrm{tr}((Im\rho_s)^2)].
\end{split}\end{equation}
When the dimension $d$ is large, we have
 \begin{equation}\label{e201}
\mathcal{E}_{o}'\approx\frac{1}{2N}[\frac{3d^2}{8}+\frac{d}{2}\left(\sqrt{\frac{d}{2}+\frac{d^2}{16}}
+1 \right)-\mathrm{tr}(\rho_s^2)].
\end{equation}
Scale MSE is used to measure the efficiency of the quantum state tomography, which is defined by
\begin{equation}\label{e21}
\mathcal{E}_s = N \mathcal{E}.
\end{equation}
In this paper we discuss QSTW in comparison with two well-established state estimation strategies. One is MUB tomography which is composed of the measurements on a complete set of mutually unbiased bases (MUBs)~\cite{qst11,qst13}. The other one is SIC tomography composed of symmetric informationally complete (SIC) measurements~\cite{qst11,qst12,qst13}.

For an unknown d-dimensional state, the optimal MSE of MUB tomography is~\cite{qst11}
\begin{equation}\label{emub}
\mathcal{E}_{s,mub}=(d+1)[d-\mathrm{tr}(\rho_s^2)].
\end{equation}
And the optimal MSE of SIC tomography is~\cite{qst11}
 \begin{equation}\label{esic}
\mathcal{E}_{s,sic}= d^2+d-1-\mathrm{tr}(\rho_s^2).
\end{equation}

By Eq. (\ref{e201}), the optimal scaled MSE of QSTW is
 \begin{equation}\label{eqstw1}
\mathcal{E}_{s,o}'\approx\frac{1}{2}[\frac{3d^2}{8}+\frac{d}{2}\left(\sqrt{\frac{d}{2}+\frac{d^2}{16}}
+1 \right)-\mathrm{tr}(\rho_s^2)].
\end{equation}
As shown in Fig. 3, the values of $\mathcal{E}_{s,o}'$ are less than the values of $\mathcal{E}_{s,mub}$ and $\mathcal{E}_{s,sic}$. It means that the efficiency of QSTW is significantly higher than the efficiencies of SIC tomography and MUB tomography with the same number of the measurements.
\begin{figure}[t]
\centering \includegraphics[scale=0.5]{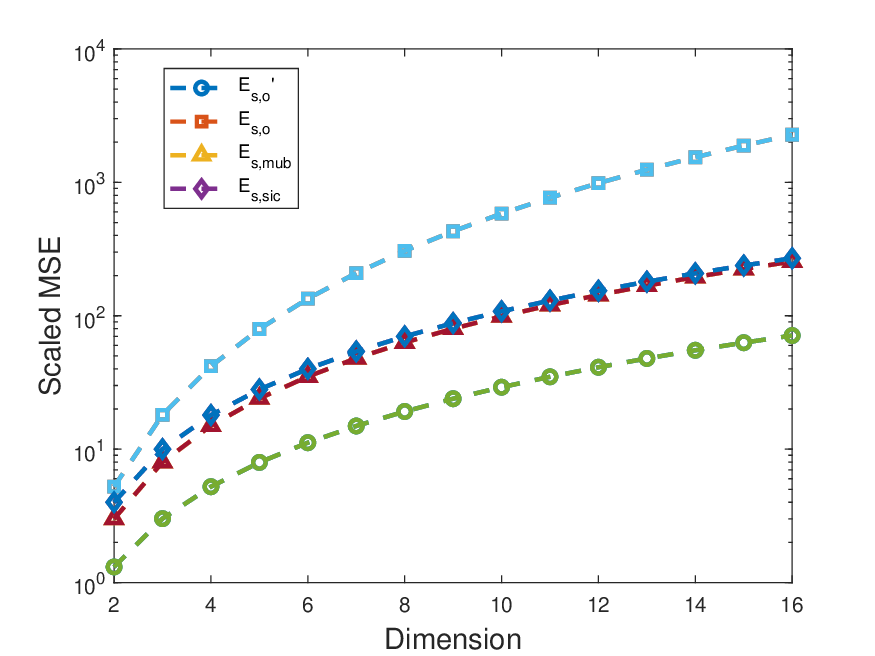} \caption{(Color online) The scaled MSEs of the different quantum tomography schemes with a pure unknown state. Circles represent $\mathcal{E}_{s,o}'$, squares represent $\mathcal{E}_{s,o}$, dianmonds represent SIC tomography, and triangles represent MUB tomography.}
\label{fig3}
\end{figure}
However, the number of the measurements is always determined by the number of the identical unknown systems in quantum state tomography. In QSTW $2dN$ systems are consumed to accomplish $N$ times measurements. It is reasonable to multiply $\mathcal{E}_o'$  by $2dN$ instead of $N$ to obtain the scaled MSE
 \begin{equation}\label{eqstw2}
\mathcal{E}_{s,o}\approx d[\frac{3d^2}{8}+\frac{d}{2}\left(\sqrt{\frac{d}{2}+\frac{d^2}{16}}
+1 \right)-\mathrm{tr}(\rho_s^2)].
\end{equation}

From Fig. 3 it appears that $\mathcal{E}_{s,o}$ is lager than $\mathcal{E}_{s,mub}$ and $\mathcal{E}_{s,sic}$. The efficiency of QSTW is less than those of SIC and MUB when the number of the measurements determined by the identically prepared systems.

\section{Conclusion}\label{sec5}
We have obtained the optimal quantum state tomography via weak value (QSTW). The optimal measurements strengths and the minimal mean squared error of the reconstructed state have been derived. The efficiency of QSTW is higher than the efficiencies of SIC and MUB with the same number of the measurements. While QSTW is less efficient than SIC and MUB when the number of the measurements determined by the identically prepared systems. The optimal QSTW discussed in this paper may be useful in reconstructing unknown quantum states.

\section*{Acknowledgments}
This work was financially supported by the National Natural Science Foundation of China (Grants No. 11965005).

\end{document}